\documentclass[11pt,a4paper]{article}
\usepackage{jheppub}
\usepackage[utf8x]{inputenc}
\usepackage{amsmath}
\usepackage{amssymb}
\usepackage{float}
\usepackage{array}
\usepackage{arydshln}
\newcolumntype{P}[1]{>{\centering\arraybackslash}p{#1}}
\numberwithin{equation}{section}

\title{\boldmath The Covariant Chiral Ring}
 \author{Antoine Bourget}
 \author{and Jan Troost}
 \affiliation{Laboratoire de Physique Théorique\footnote{Unité Mixte du CNRS et de l'École Normale Supérieure associée à l'université Pierre et Marie Curie 6, UMR 8549.}, \\
 École Normale Supérieure, \\
 24 rue Lhomond, 75005 Paris, France}
 
\emailAdd{bourget@lpt.ens.fr}
\emailAdd{troost@lpt.ens.fr}

\abstract{We construct a covariant generating function for the spectrum of chiral primaries of symmetric orbifold conformal field theories with $N=(4,4)$ supersymmetry in two dimensions. For seed target spaces $K3$ and $T^4$, the  generating functions capture the $SO(21)$ and $SO(5)$ representation theoretic
content of the chiral ring respectively.  Via string dualities, we relate the transformation properties of the chiral ring
under these isometries of the moduli space to the Lorentz covariance of perturbative string partition functions in flat space. }

\begin{document} 
\maketitle
\flushbottom

\section{Introduction}

The AdS/CFT duality \cite{Maldacena:1997re} between two-dimensional 
conformal field theories and gravitational theories in three-dimensional
anti-de Sitter space has been
a powerful tool for studying the microscopic origin of black hole
entropy \cite{Strominger:1996sh,Strominger:1997eq}. It is a testing ground for holography that is
under  good control, at both small and large curvature. For instance, protected three-point functions 
can be computed exactly as
a function of $\alpha'$ over the radius of curvature in the bulk
string theory \cite{Gaberdiel:2007vu,Dabholkar:2007ey}, and can be matched to the
three-point functions of the boundary theory \cite{Jevicki:1998bm,Lunin:2001pw}.
Moreover, we can calculate all $\alpha'$ corrections to the central charge, 
without relying on supersymmetry \cite{Troost:2011ud}.

The three-point functions provided a dynamical
test of the holographic correspondence, albeit in a limited sense. In the case where
the boundary theory exhibits an $N=(4,4)$ supersymmetry,
it has been proven that certain three-point functions
satisfy a non-renormalisation theorem. They are covariantly constant on the moduli
space of the superconformal field theory \cite{deBoer:2008ss}.  Consequently, the vector bundle of chiral
primaries over the symmetric space of moduli is homogeneous. The bundle is then determined 
completely by the representation of the holonomy group constituted by the fibres.

In the example of IIB superstring theory on $AdS_3 \times S^3 \times
M$, where $M$ is either  the manifold $K3$ or the four-torus $T^4$, the
three-point functions have been computed on both sides of the duality,
and successfully compared \cite{Gaberdiel:2007vu,Dabholkar:2007ey}. In
particular, these three-point functions have been calculated in the bulk at a point
of moduli space where the theory is described by $AdS_3 \times
S^3 \times M$ with only a background NSNS flux, corresponding to $N_1$
fundamental strings and $N_5$ NS5-branes. On the boundary, the
three-point functions have been computed at the symmetric orbifold
point corresponding to the conformal field theory on $Sym_{N_1 N_5}
(M)$. The identification of the conformal field theory is inspired
by the physics of the $S$-dual D1-D5 Higgs branch (but the identification
remains subtle \cite{Seiberg:1999xz}).
The comparison of three-point functions has been successfully carried out at leading order
in $1/N$ where $N= N_1 N_5$.

In \cite{Dijkgraaf:1998gf,deBoer:2008ss}, it was
argued that the full covariance of the chiral ring of these models is
$SO(21)$ for $M=K3$ and $SO(5)$ for $M=T^4$. Moreover, the $SO(n)$ covariant
representation content of the chiral ring was identified for operators
with small R-charges
for the case $M=K3$. In this paper, we continue this line of reasoning as follows.
We briefly review the proposed covariance in section \ref{properties}.
Motivated by the importance of the representation theoretic content not
only for a more refined classification of the spectrum, but also for the 
protected three-point functions,
 we wish to identify the $SO(21)$ and $SO(5)$
representations appearing in the chiral ring at all levels. To
that end, we will construct a fully covariant generating function
in section \ref{covariant}, both for the case of $M=K3$ and $M=T^4$.

\section{The Chiral Ring}
\label{properties}
In this section, we review salient features of the background of type IIB string theory under study.
We consider type IIB string theory compactified on $M=T^4$ or $M=K3$, and
a string transverse to $M$, consisting of D1-strings and D5-branes wrapped on $M$. 
We will study the background in the vicinity of the string.
The near-string geometry
is $AdS_3 \times S^3 \times M$. The bulk string theory is dual \cite{Maldacena:1997re} to a
two-dimensional $N=(4,4)$ superconformal field theory on the boundary of level
$N=N_1 N_5$.
The moduli space of these theories is locally of the form
\begin{equation}
 \frac{SO(4,n)}{SO(4) \times SO(n)} \, , 
\end{equation}
where $n=21$  and $n=5$ for $K3$ and $T^4$ respectively \cite{Dijkgraaf:1998gf}. 
It is generally assumed that there is a point in the moduli space where the conformal field theory can be described as a 
symmetric product conformal field theory $Sym_N (M)$ where $N=N_1 N_5$.

The chiral ring of the theory with respect to a $N=(2,2)$ superconformal
subalgebra
forms a vector bundle over the moduli space.
The symmetry group of the bundle of chiral primaries is
$SO(4) \times SO(n)$.
The three-point functions are covariantly constant on the moduli space \cite{deBoer:2008ss},
and the curvature of the bundle of chiral primaries is covariantly constant as well \cite{deBoer:2008ss}.
Thus the chiral primaries give rise to a homogeneous bundle on a symmetric
space whose geometry is determined in terms of the representation of the
fibre with respect to the structure group $SO(4) \times SO(n)$ and
the connection of the tangent bundle of the symmetric space. The curvature of the 
tangent bundle can be computed in conformal field theory, and turns out to be trivial
in the $SO(4)$ factor of the structure group \cite{deBoer:2008ss}. Thus, it is sufficient to characterize
the $SO(n)$ representations of the chiral primaries to fully characterize the connection
of the chiral primary vector bundle on the moduli space \cite{deBoer:2008ss}. 

Beyond the elementary observation that
we obtain knowledge of the spectrum by classifying it with respect to the largest 
symmetry group of the problem, we believe that the fact that the enhanced symmetry also 
provides constraints on 
the three-point functions of the theory are good motivations for explicitly identifying the full
representation theory content of the chiral ring. To that end, we will construct a fully covariant
generating function.

\section{The Covariant Generating Function}
\label{covariant}
In this section, we firstly  review the generating function of the  ring
of left and right chiral primaries. In a $N=(4,4)$ supersymmetric
theory, the $(c,c)$ ring with respect to a given $N=(2,2)$ superconformal
subalgebra of the $N=(4,4)$ superconformal algebra is isomorphic to its $(a,c)$, $(c,a)$ and $(a,a)$ anti-chiral 
counterparts. It is therefore sufficient to consider the $(c,c)$ chiral ring only.
\subsection{The counting function}
Our starting point is the generating function for the chiral ring
of the superconformal field theory with target space $M$, which in a geometric regime
can be identified as the Poincar\'e polynomial $P_{t,\bar{t}}$ of the manifold $M$.
We can alternatively view it as the trace over the $NSNS$ Hilbert space, with
states weighted by their left and right $U(1)_R$ charges $(2 J_0, 2 \bar{J}_0)$, 
or as the generating function
for the Hodge numbers $h_{p,q}$:
\begin{eqnarray}
P_{t,\bar{t}}(M) &=& Tr_{NSNS} (t^{2 J_0} \bar{t}^{2 \bar{J}_0})
\nonumber \\
&=& \sum_{p,r=0}^{c/3} h_{p,r} t^p \bar{t}^r \, .
\end{eqnarray}
Such a polynomial will be represented using the associated Hodge diamond in the following way:
\begin{equation}
\sum\limits_{p,r=0}^m h_{p,r} t^p \bar{t}^r =  
\begin{array}{ccccccccc}
 & & & & h_{0,0} & & & & \\
& & & h_{1,0} & & h_{0,1} & & & \\
 & & h_{2,0} & & h_{1,1} & & h_{0,2} &&  \\
& \dots & & \dots & & \dots & & \dots &  \\
h_{m,0}&  &  \dots&  & \dots &  & \dots&  & h_{0,m}\\
& \dots & & \dots & & \dots & & \dots&  \\
&  & h_{m,m-2} & & h_{m-1,m-1} & & h_{m-2,m} & & \\
&  & & h_{m,m-1} & & h_{m-1,m} & & & \\
 &  & & & h_{m,m} & & && 
\end{array}
\end{equation}
where $m$ is the complex dimension of $M$, and we delete the coefficients $h_{p,q}$ where $p+q$ is odd when these coefficients identically vanish (as for instance in diamond (\ref{HodgeK3})). Moreover we have the relations $h_{p,r}=h_{r,p}$ from complex conjugation, $h_{p,r}=h_{m-p,m-r}$ from Poincar\'e duality and $h_{p,r}=h_{m-p,r}$ if $M$ is hyperk\"ahler. We then sometimes represent only the upper-left eighth of the diamond: 
\begin{equation}
\label{8thdiamond}
\sum\limits_{p,r=0}^m h_{p,r} t^p \bar{t}^r =  
\begin{array}{ccccc}
& & & & h_{0,0} \\
& & & h_{1,0} &  \\
&  & \dots & & h_{1,1} \\
  & \dots & & \dots & \\
 h_{\frac{m}{2},0} & & \dots & &  \dots \\
  & \dots & & \dots & \\
&  & \dots & & h_{\frac{m}{2}-1,\frac{m}{2}-1} \\
&  & & h_{\frac{m}{2},\frac{m}{2}-1} &  \\
 &  & & & h_{\frac{m}{2},\frac{m}{2}} \\
\end{array}
\end{equation}
The generating function $Z$ for the ring of the symmetric product theory can be written down in terms
of the data of the underlying theory:
\begin{eqnarray} \label{FuncSym}
 Z(M^N/S_N)   &=& \sum\limits_{N \geq 0} q^N P_{t , \bar{t}} (M^N /S_N) \nonumber \\
  &=& \prod\limits_{m=1}^{+
\infty} \prod\limits_{p,r=0}^{c/3} \left( 1+ (-1)^{p+r+1} q^m t^{m+p-1}
\bar{t}^{m+r-1} \right)^{(-1)^{p+r+1} h_{p,r}} \, .
 \end{eqnarray}
The formula has a physical interpretation in terms of a second quantized string
theory, with twisted long string sectors \cite{Dijkgraaf:1996xw}. The new variable $q$ keeps
track of the number of copies of the manifold $M$ that are in play.
Note that the R-charge is augmented by $m-1$ both on the left
and the right, for each extra oscillator that comes in at order $m$.
To make further progress in the analysis of the covariance of the generating function,
we consider the two specific examples of $K3$ and $T^4$ separately, in sections
\ref{sectionK3} and \ref{sectionT4}.

\subsection{Type IIB string theory on $AdS_3 \times S^3 \times K3$}
\label{sectionK3}
We first review  properties of the chiral ring partition function of $K3$
and its second quantized interpretation. We then present the covariant generating function.
Finally, we relate the latter  to the Lorentz covariant partition function of the chiral,
massive and bosonic half of the heterotic string.
\subsubsection{The chiral ring partition function}
The only input we need to render the chiral ring counting function $Z$ in equation (\ref{FuncSym}) explicit is the
Hodge diamond of $M=K3$:
\begin{equation}
P_{t , \bar{t}} (K3) =  \begin{array}{ccccc}
  & & 1 & &  \\
& 0 & & 0 & \\
  1 & & 20 & & 1 \\
& 0 & & 0 & \\
&  & 1 & &
\end{array}
= 
 \begin{array}{ccc}
   & 1 &   \\
  1 &  20 &  1 \\
  & 1 & 
\end{array}
\label{HodgeK3}
\end{equation}
Since the Hodge numbers are only non-zero for $p+q$ even, the second quantized
string theory is bosonic (for one chirality) and the partition function
consists of a multiplication of denominators:
\begin{equation}
 Z(K3^N/S_N) = \sum\limits_{N \geq 0} q^N P_{t , \bar{t}} (K3^N /S_N) = \prod\limits_{m=1}^{+
\infty} \frac{1}{\mathfrak{P}_m (q,t,\bar{t})} \, ,
 \end{equation}
where the factors are determined by the entries of the Hodge diamond:
\begin{eqnarray*}
 \mathfrak{P}_m (q,t,\bar{t}) = 
 \begin{array}{ccc}
  & \left( 1- q^m t^{m-1} \bar{t}^{m-1} \right) \times &  \\
  \left( 1- q^m t^{m+1} \bar{t}^{m-1} \right) & \left( 1- q^m t^{m} \bar{t}^{m} \right)^{20} & \left( 1- q^m t^{m-1} \bar{t}^{m+1} \right)   \times \\
  &  \left( 1- q^m t^{m+1} \bar{t}^{m+1} \right) \, .   & 
 \end{array}
\label{vectordecomp}
\end{eqnarray*}
Using this generating function,
we can compute the number
of chiral primaries as a function of R-charges, for any given $N$.
As an elementary example, we quote the informative eighth of the diamond for the case of $N=8$:
\begin{equation}
\begin{array}{P{1.6cm}P{1.6cm}P{1.6cm}P{1.6cm}P{1.6cm}P{1.6cm}P{1.6cm}P{1.6cm}P{1.6cm}P{1.6cm}}
  &   &   &   & 1 \\
&   &   & 1 &   21 \\
 &   & 1 &    22 &    254 \\
  & 1    & 22    & 276    & 2278 \\
 1    & 22    & 277    & 2553    & 16744 \\
 & 277    & 2575    & 19274    & 106284 \\
  &   & 19528   & 125006   & 599470 \\
 &   &   & 702926   & 2983928 \\
  &   &   &   & 11251487 \\
\end{array}
\end{equation}
In this table, we observe for instance that the Hodge number $h_{1,1}=20$ for one copy of $K3$ has been
augmented to $h_{1,1}=21$ for any number of copies of $K3$ larger than one. 
A given coefficient $h_{r,s}$ can only receive contributions from
polynomials $ \mathfrak{P}_m (q,t,\bar{t}) $ with $m \leq r+s$. Thus, the entries
of the table stabilise at $m = r+s$. As we increase the number of copies $N$ of the seed manifold $M$, the
entries will no longer change.

\subsubsection{A second quantized perspective}
 A convenient way
to represent the operators in the chiral ring is found in terms of the second quantized view on the generating
function of the elliptic genus \cite{Dijkgraaf:1996xw}.
All the chiral primaries can be represented in terms of the set of operators of the original
theory and the twisted sector ground states (labeled by $m$). More explicitly, we can define operators:
\begin{equation}
 \alpha_{-m}^{(p,q),i} 
\end{equation}
which live in a conformal field theory corresponding to $m$ copies
of the original theory, and has left and right R-charges $(m+p-1,m+q-1)$.
The label $i$ represents the multiplicity of the $(p,q)$ entry in the 
original Hodge diamond.
For $M=K3$ we have that $i=1,2,\dots,20$ for $(p,q)=(1,1)$, and the index
can be dropped for other values of $(p,q)$. For $K3$, all oscillators are bosonic.
The generating function given above can then
 be interpreted as the partition function of a string theory with the
following list of possible excitations:
\begin{table}[H]
\centering
\begin{tabular}{|c|c|c|}
\hline
Name & Number  & R-Charges  \\
\hline
$\alpha_{-n}^{(0,0)}$ & 1 & $(n-1,n-1)$ \\
\hline
$\alpha_{-n}^{(2,0)}$ & 1 & $ (n+1,n-1)$ \\
\hline
$\alpha_{-n}^{(1,1),i}$ & 20 & $ (n,n)$ \\
\hline
$\alpha_{-n}^{(0,2)}$ & 1 & $ (n-1,n+1)$ \\
\hline
$\alpha_{-n}^{(2,2)}$ & 1 & $ (n+1,n+1)$ \\
\hline
\end{tabular}
\caption{The list of excitations of the second quantized chiral string.}
\label{ListOfK3Excitations}
\end{table}
\noindent
Let us make more explicit lists of 
low lying chiral primaries \cite{deBoer:2008ss}. The number of copies of the original theory
has to be equal to the total number of copies $N$. We take $N$ to be sufficiently
large for the Hodge numbers of the orbifold to have stabilized. If necessary,
we can always add $\alpha_{-1}^{(0,0)}$ to find the total number $N$ without changing
any of the other charges of the operator. In accord with the Hodge diamond entries
recorded before, we find the table:
\begin{table}[H]
\centering
\begin{tabular}{|c|c|c|c|c|c|c|}
\hline
R-charge & Number of states  & States  \\
\hline
$(0,0)$ & 1 & $ | N \rangle \equiv (\alpha_{-1}^{(0,0)})^N | 0
\rangle $ \\
\hline
$(2,0)$ & 1 & $\alpha_{-1}^{(2,0)} |N-1 \rangle$ \\
\hline
$(1,1)$ & 21 & $\alpha_{-2}^{(0,0)} |N-2 \rangle$ and $\alpha_{-1}^{(1,1),i} |N-1 \rangle$ \\
\hline
$(0,2)$ & 1 & $\alpha_{-1}^{(0,2)} |N-1 \rangle$ \\
\hline
$(4,0)$ & 1 & $(\alpha_{-1}^{(2,0)})^2 |N-2 \rangle$ \\
\hline
$(3,1)$ & 22 & $\alpha_{-1}^{(2,0)} \alpha_{-1}^{(1,1),i} |N-2 \rangle$ , $\alpha_{-1}^{(2,0)}
\alpha_{-2}^{(0,0)} |N-3 \rangle$ , $\alpha_{-2}^{(2,0)} |N-2 \rangle$ \\
\hline
$(2,2)$ & 254 & $\alpha_{-1}^{(1,1),i} \alpha_{-2}^{(0,0)}|N-3 \rangle$ , $\alpha_{-1}^{(1,1),i}
\alpha_{-1}^{(1,1),j}|N-2 \rangle$ , \\
 & & $\alpha_{-3}^{(0,0)}|N-3 \rangle $ , $\alpha_{-2}^{(1,1),i}|N-2 \rangle$ ,  \\
  & & $\left( \alpha_{-2}^{(0,0)} \right)^2 |N-4 \rangle$ , $\alpha_{-1}^{(2,0)} \alpha_{-1}^{(0,2)}|N-2 \rangle $, $ \alpha_{-1}^{(2,2)}|N-1 \rangle $ \\
  \hline
\end{tabular}
\caption{The list of states with given (left,right) R-charges.
}
\label{K3table}
\end{table}
\vspace{3mm}
\noindent
Our goal is to derive a generating function that organizes the chiral primaries which manifestly form
representations of $SO(20)$, into representations of $SO(21)$. We wish to find a generating function
at all levels, thus generalizing the 
analysis at low levels performed in \cite{deBoer:2008ss}. 

\subsubsection{The covariant generating function}
\label{covK3}
There is a manifest $SO(20)$ symmetry in the original K3 Hodge diamond (\ref{HodgeK3}) that
rotates the vector of $(1,1)$ harmonic forms, and otherwise
acts trivially. This $SO(20)$ action is inherited by the symmetric product
theory. To render the $SO(21)$ representation content of the 
generating function manifest, we follow a slightly roundabout route whose 
logic will become apparent in due course.
Firstly, we write a  bosonic chiral string partition function $Z_{24}$ with
$SO(24)$ symmetry. The fugacities are
$y_i$ where $i=1,2, \dots,12$ for the vector representation
of $SO(24)$:\footnote{We refer to appendix \ref{representationtheory} for the relevant representation
theory and nomenclature.}
 \begin{equation}
Z_{24} =   \prod\limits_{m=1}^{\infty} \prod\limits_{\lambda \in \Lambda_{\mathbf{24}}} \left( 1- q^m
 \prod\limits_{i=1}^{12} y_i^{\lambda_i} \right)^{-1} \, .
 \end{equation}
 We introduced the notations $\lambda$ for the weights inside the set of weights $\Lambda_{\mathbf{24}}$
 of the $24$-dimensional vector representation of $SO(24)$, as well as their components $\lambda_i$ in a basis
 of fundamental weights.
It is important to realize at this stage that we know that this partition function is, in fact, $SO(25)$
covariant starting at level $q^2$, since the massive modes of a bosonic open string (or the non-supersymmetric
chirality of a heterotic string) exhibit the covariance
of the little group $SO(25)$ in $25+1$ dimensions. This can be made manifest by defining the $SO(25)$ covariant
version which is obtained after the substitution (see equation (\ref{substitution})) \cite{Hanany:2010da}:
\begin{equation}
y_{11} = z_{11}/z_{12} \, ,
\end{equation}
and renaming the other variables $y_i=z_i$ (for $1 \leq i \leq 12$, $i \neq 11$), where the variables $z_i$ are $SO(25)$ fugacities.
As an example, let us illustrate how this substitution reconstitutes the character
of the $SO(25)$ vector representation from the vector of $SO(24)$ and a singlet: starting from
\begin{equation}
\chi_{\mathbf{24}} = y_1 + \frac{1}{y_1} + \left[ \sum\limits_{i=1}^9 \frac{y_i}{y_{i+1}} +  \frac{y_{i+1}}{y_{i}} \right] + \frac{y_{10}}{y_{11} y_{12}} +  \frac{y_{11} y_{12}}{y_{10}} + \frac{y_{11}}{y_{12}} + \frac{y_{12}}{y_{11}}
\end{equation}
we obtain after substitution of the new variables 
\begin{equation}
1+ \chi_{\mathbf{24}} \rightarrow 1+ z_1 + \frac{1}{z_1} + \left[ \sum\limits_{i=1}^{10} \frac{z_i}{z_{i+1}} +  \frac{z_{i+1}}{z_{i}} \right] + \frac{z_{11}}{z_{12}^2} + \frac{z_{12}^2}{z_{11}} = \chi_{\mathbf{25}} \, . 
\end{equation}
The underlying mechanics of these substitution rules is the equivalence of the covariant BRST
cohomology (with Lorentz invariant ghosts and Lorentz covariant matter) with the light cone
gauge spectrum.
 This $SO(25)$ covariant partition function (at level $2$ and higher) \cite{Hanany:2010da} can be fruitfully adapted to the present context. Let us proceed formally at this stage. We first of all assign both left and right $R$-charges equal to the string mass squared, or level, through the substitution $q \rightarrow t \bar{t} q$.
We furthermore fine-tune the left and right R-charge assignment using the charges
of the states under a $U(1) \times U(1) \subset SO(4)$ subgroup of  $SO(4) \times SO(21) \subset SO(25)$. We can implement this through the further substitution:
\begin{equation}
 \left\{
 \begin{array}{l}
z_i \longrightarrow \tilde{z}_i    \qquad   \mathrm{for} \;   i=1,...,8\\
  z_9 \longrightarrow t \bar{t} \tilde{z}_8 \\
  z_{10} \longrightarrow t^2 \tilde{z}_8 \\
  z_{11} \longrightarrow t^2  \tilde{z}_9 \\
  z_{12} \longrightarrow t  \tilde{z}_{10} \, .
 \end{array}
 \right.
\end{equation}
 Following the scheme outlined above, we find the $SO(21)$ covariant generating function -- we drop the tildes
on the variables $\tilde{z}_i$  in the final result --:
 \begin{eqnarray}
 \label{covariantK3genfun}
 \tilde{P}(q,t,\bar{t},z_1,...,z_{10}) 
&=& \prod\limits_{m=1}^{+
\infty}  \frac{\left( 1- q^m
t^{m} \bar{t}^{m}\right)}{\prod\limits_{\lambda' \in
\Lambda_{\mathbf{21}}} \left( 1- q^m
t^{m} \bar{t}^{m} \prod\limits_{i=1}^{10} z_i^{\lambda'_i} \right)   } \\ &  & \mkern-130mu \times \prod\limits_{m=1}^{+
\infty} \frac{1 }{
\left( 1- q^m t^{m-1} \bar{t}^{m-1} \right)
\left( 1- q^m t^{m+1} \bar{t}^{m-1} \right) 
\left( 1- q^m t^{m-1} \bar{t}^{m+1} \right) 
\left( 1- q^m t^{m+1} \bar{t}^{m+1} \right) } \nonumber
 \end{eqnarray}
In the final expression, the coefficient of the monomial $t^p \bar{t}^q$ is the character of the
representation of $SO(21)$ in which the chiral primaries of left and right R-charges
$(p,q)$ transform.
As an example we can study the multiplicity coefficient 254 for $(p,q)=(2,2)$. We have to expand to the fourth
power in $q$
to find it. Using a symbolic manipulation program, we recognize the
character of the representation
$\mathbf{230}+\mathbf{21}+3\times \mathbf{1}$ of $SO(21)$. In table \ref{K3table}, the corresponding chiral primaries have been organized in three lines according to this decomposition. 
The coefficient 2278 of the term with $(p,q)=(3,3)$ corresponds to the
$\mathbf{1750}+\mathbf{230}+\mathbf{210}+4 \times \mathbf{21} + 4 \times
\mathbf{1}$ of $SO(21)$.
As a final example, we can decompose the coefficient 16744 at $(p,q)=(4,4)$
in representations of $SO(21)$ as 
\begin{equation}
\mathbf{16744} = 
\mathbf{10395} +\mathbf{3059} +\mathbf{1750} + 5 \times \mathbf{230} + \mathbf{210}
+8 \times \mathbf{21} +12 \times \mathbf{1} \, .
\end{equation}
As we mentioned, and as is apparent from the numerator in formula (\ref{covariantK3genfun}), we  made
use of the relation between
the chiral ring generating function, and the covariantized generating function of
24 chiral bosons, as embedded in the covariant bosonic or heterotic
string (through the addition of light-cone oscillators as well as ghosts  \cite{Hanany:2010da}).
Let us finally show how known dualities link up these ideas, and provide method to this madness.

\subsubsection{The duality}
If we return to our starting point, which was a
D1-D5 system on $\mathbb{R}^{5,1} \times K3$, and we compactify on an extra circle along the string, then
we can dualize the configuration to a heterotic string
with momentum $n=N_1$ and winding $w=N_5$ on that circle.\footnote{An example duality chain would be a chain of  first
$S$-duality,
then $T$-duality along the circle, and then IIA/Heterotic duality on $K3/T^4$.} The elliptic genus
of the conformal field theory dual to $AdS_3 \times S^3$
 is indeed equal to the counting function
of half BPS states in heterotic string theory on $S^1 \times T^4$ in asymptotically flat space, with
momenta $p_{L,R} = n/R \pm w R /\alpha'$ on the circle $S^1$ and no momentum
on the $T^4$. See e.g.  \cite{Vafa:1994tf}
for a foreshadowing of this known fact. The counting is mostly unaffected by the procedure of taking the 
near-brane limit. The states that are counted preserve half the supersymmetry, are in one of
sixteen right-moving ground states, and obey the equations \cite{Dabholkar:1990yf}:
\begin{eqnarray}
\frac{\alpha' m^2}{4} &=& \frac{\alpha'}{4} p_R^2 
\nonumber \\
&=& \frac{\alpha'}{4} p_L^2 + N_L -1 \, ,
\end{eqnarray}
where the mass $m$ is measured transversely to the circle, and 
$N_L$ is the left-moving oscillator number. For the left-movers, we 
allow any of the 24 bosonic oscillator excitations (but no further
zero mode excitations). The oscillator level is restricted
by level matching to satisfy:
\begin{eqnarray}
N_L -1 &=& -nw \, .
\end{eqnarray}
We note that for non-zero quantum numbers $(n,w)$, we are dealing with massive
states only.
We thus find that only the bosonic half of the heterotic string, and only the oscillator modes
intervene in the half BPS state counting
\cite{Dabholkar:1990yf}. Note that the oscillator states are
as in the chiral half of a $25+1$ dimensional bosonic string theory, compactified on
$T^4$.
The little group for massive string states in this space is $SO(21)$ -- we can safely
ignore the compactification on the extra $S^1$, by realizing
that we have fixed the momentum in this direction, and that in regard to state counting, the direction is 
otherwise equivalent
to a non-compact direction in space. Thus, the chiral ring spectrum indeed naturally permits
an action of $SO(21)$.

We can thus follow \cite{Hanany:2010da} and write down the covariant partition function. To perform this final step,
we first discuss how to read off the R-charges of the states from the chiral heterotic string
perspective. The R-charge of a state in the dual conformal field theory is given by:
\begin{eqnarray}
(p,q) &=& (N_{L},N_{L}) + (p^D,p^A)
\end{eqnarray}
where $(p^D,p^A)$ are generators of a diagonal and anti-diagonal
$SO(2)$ inside the $SO(4)$ action on the tangent space to the
four-torus $T^4$.  The oscillators at given level transform in the
$\mathbf{4}$ of $SO(4)$ and will carry charges $(\pm 1, \pm 1)$ under
the $SO(2) \times SO(2)$ subgroup. These charges dictate the fugacity substitution rules we
declared above.
To complete the proof of $SO(21)$
covariance of the generating function from this dual perspective, 
we firstly remark that we can compute the covariant BRST
cohomology at each fixed oscillator number $N_L$. And the second and final argument is
that $SO(21)$ commutes with the $SO(4)$ acting on the tangent
space to $T^4$.  

Finally, we note that our proof is also constructive in providing
generators of $SO(21)$. The generators are the
right-moving and massive part of the $SO(21)$ Lorentz generators in light-cone gauge. Their
explicit expression is therefore known (see e.g. \cite{GSW1}). This provides
a hands-on construction of the generators, and thus answers 
a question raised in  \cite{deBoer:2008ss}, in this duality frame, at a particular point in the 
moduli space.

In summary, through
string-string duality, we have related the $SO(21)$ covariance of the chiral ring of the superconformal field
theory dual to $AdS_3 \times S^3 \times T^4$ to 
Lorentz covariance of the heterotic string on $T^4$. This relation allowed us to construct the covariant
generating function for the chiral ring.

\subsection{Type IIB string theory on $AdS_3 \times S^3 \times T^4$}
\label{sectionT4}
We wish to apply a similar logic to obtain the covariant generating function for the
chiral ring of the $N=(4,4)$ conformal field theory dual to Type IIB
compactified on the manifold $M=T^4$.  In this case, the scalar moduli parametrize the coset
$SO(5,5)/(SO(5) \times SO(5))$.  After considering a $D1/D5$ system, and taking
the near-brane limit, one finds that in the  $AdS_3 \times S^3 \times T^4$ geometry,
five scalars become massive (namely, the volume of $T^4$, the three
components of the anti-self-dual part of the Neveu-Schwarz two-form, and a linear
combination of the Ramond-Ramond scalar and four-form).  The twenty remaining scalars
parametrize the coset $SO(4,5)/(SO(4) \times SO(5))$. This moduli
space is again a homogeneous space.  The holonomy of the tangent
bundle of the moduli space is $SO(4) \times SO(5)$. As explained in
\cite{deBoer:2008ss}, each of the vector bundles of chiral
primaries is characterized by a representation of $SO(5)$.

The seed theory for the dual  $N=(4,4)$ superconformal field theory is the theory
on the four-torus.
The chiral ring of the seed is determined by the Hodge diamond of $T^4$: 
\begin{equation}
P_{t,\bar{t}}(T^4)=
\begin{array}{ccccc}
   &   & 1 &   &   \\
   & 2 &   & 2 &   \\
 1 &   & 4 &   & 1 \, .\\
   & 2 &   & 2 &   \\
   &   & 1 &   &   \\
\end{array}
\label{HodgeT4}
\end{equation}
The Hodge diamonds for $(T^4)^N/S_N$ are given by formula 
(\ref{FuncSym}). From this we can write the partition function for the number of chiral primaries 
at given $N$, that starts out as
\begin{equation}
\label{CFTpartfunc}
Z((T^4)^N/S_N) = 1+16 q+144 q^2+960 q^3+5264 q^4+25056 q^5+... \, ,
\end{equation}
and we can  generate the upper-left
eighth of the Hodge diamond 
for, say, the $N=9$ symmetric product:
\begin{equation}
\label{HodgeDiamondT4}
P_{t,\bar{t}}((T^4)^9/S_9) = {\scriptsize
\begin{array}
{P{1cm}P{1cm}P{1cm}P{1cm}P{1cm}P{1cm}P{1cm}P{1cm}P{1cm}P{1cm}}
   &   &   &   &   &   &   &   &   & 1 \\
   &   &   &   &   &   &   &   & 2 &   \\
   &   &   &   &   &   &   & 2 &   & 9 \\
   &   &   &   &   &   & 2 &   & 18 &   \\
   &   &   &   &   & 2 &   & 23 &   & 61 \\
   &   &   &   & 2 &   & 24 &   & 116 &   \\
   &   &   & 2 &   & 24 &   & 154 &   & 327 \\
   &   & 2 &   & 24 &   & 170 &   & 600 &   \\
   & 2 &   & 24 &   & 175 &   & 817 &   & 1514 \\
 2 &   & 24 &   & 176 &   & 932 &   & 2690 &   \\
   & 24 &   & 176 &   & 978 &   & 3695 &   & 6244 \\
   &   & 176 &   & 994 &   & 4306 &   & 10774 &   \\
   &   &   & 998 &   & 4587 &   & 14770 &   & 23278 \\
   &   &   &   & 4664 &   & 17224 &   & 38444 &   \\
   &   &   &   &   & 18010 &   & 50637 &   & 75320 \\
   &   &   &   &   &   & 55226 &   & 111912 &   \\
   &   &   &   &   &   &   & 127496 &   & 181359 \\
   &   &   &   &   &   &   &   & 213214 &   \\
   &   &   &   &   &   &   &   &   & 253508 \\
\end{array}
}
\end{equation}
\noindent
In the symmetric product conformal field theory, we can factor out the center of mass degree of freedom, which is one copy of $T^4$. This translates into the fact that the Poincar\'e polynomial $P_{t,\bar{t}}(T^4)$ divides $P_{t,\bar{t}}((T^4)^N/S_N)$. Let us define
\begin{equation}
Q_{N} (t,\bar{t} )= \frac{P_{t,\bar{t}}((T^4)^N/S_N)}{P_{t,\bar{t}}(T^4)} \in \mathbb{Z}[t,\bar{t}] \, . 
\end{equation}
\noindent
For reference, let us also record the corresponding Hodge diamond, again for $N=9$:
\begin{equation}
Q_{9} (t,\bar{t} ) = 
\label{reducedHodgeT4}
\begin{array}
{P{0.95cm}P{0.95cm}P{0.95cm}P{0.95cm}P{0.95cm}P{0.95cm}P{0.95cm}P{0.95cm}P{0.95cm}}
   &   &   &   &   &   &   &   & 1 \\
   &   &   &   &   &   &   & 0 &   \\
   &   &   &   &   &   & 1 &   & 5 \\
   &   &   &   &   & 0 &   & 4 &   \\
   &   &   &   & 1 &   & 6 &   & 22 \\
   &   &   & 0 &   & 4 &   & 28 &   \\
   &   & 1 &   & 6 &   & 34 &   & 94 \\
   & 0 &   & 4 &   & 32 &   & 140 &   \\
 1 &   & 6 &   & 35 &   & 167 &   & 376 \\
   & 4 &   & 32 &   & 172 &   & 588 &   \\
   &   & 35 &   & 179 &   & 718 &   & 1400 \\
   &   &   & 176 &   & 768 &   & 2200 &   \\
   &   &   &   & 787 &   & 2696 &   & 4714 \\
   &   &   &   &   & 2844 &   & 7032 &   \\
   &   &   &   &   &   & 7946 &   & 12744 \\
   &   &   &   &   &   &   & 15508 &   \\
   &   &   &   &   &   &   &   & 19617 \\
\end{array}
\end{equation}
\noindent
It may be useful to stress the fact that three different $SO(4)$ symmetries are at play. One is the R-symmetry $SO(4)^R = SU(2)^R_L \times SU(2)^R_R \supset U(1)^R_L \times U(1)^R_R$, generated by the $3+3$ bosonic currents of the superconformal algebra, which rotates both these currents and the supercurrents. Another is the outer automorphism group $SO(4)^{outer}$ under which only the 4 supercurrents transform. Finally, there is the transformation group $SO(4)^{T^4}=SU(2)^{T^4}_L \times SU(2)^{T^4}_R$ of the tangent space which rotates the two complexified bosons and fermions. A given chiral primary field of the seed $T^4$ conformal field theory
with $U(1)^R_L \times U(1)^R_R$ charge $(p,q)$ is a singlet of $SO(4)^{outer}$, it is the highest weight state of the spin $p/2$ (respectively $q/2$) representation of $SU(2)^R_L$ (respectively $SU(2)^R_R$), and belongs to the following representations of $SU(2)^{T^4}_L \times SU(2)^{T^4}_R$: 
\begin{equation}
\label{HodgeT4rep}
\begin{array}{ccccc}
   &   & (0,0) &   &   \\
   & (\frac{1}{2},0) &   & (0,\frac{1}{2}) &   \\
 (0,0) &   & (\frac{1}{2},\frac{1}{2}) &   & (0,0) \\
   & (\frac{1}{2},0) &   & (0,\frac{1}{2}) &   \\
   &   & (0,0) &   &   \\
\end{array}
\end{equation}
It is the latter property that will be most useful in the following, since we will embed the $SO(4)^{T^4}$ acting on the tangent space of $T^4$ into the Lorentz group of flat space.
Combining these representations (\ref{HodgeT4rep}) with the formula (\ref{FuncSym}) where the product is to be interpreted as tensor product of representations, we readily obtain the representations of $SU(2)^{T^4}_L \times SU(2)^{T^4}_R$
in which the chiral primaries of (\ref{HodgeDiamondT4}) transform. For instance, the $2$ on the upper line transforms as a $(\frac{1}{2},0)$ for odd $p$ and as $2 \cdot (0,0)$ for even $p$, while the $9$ decomposes as $2 \cdot (\frac{1}{2},\frac{1}{2}) + (0,0)$. 

\subsubsection{The covariant generating function}
String-string duality  now  motivates the attempt to 
implement the $SO(5)$ covariance of the chiral ring generating function by exploiting the $SO(9)$ covariance
of the massive superstring spectrum in flat space. This time around, we will use the
$SO(9)$ covariance of the left chiral half of the type II
superstring. A hands-on way to motivate this is to observe that
the Hodge diamond (\ref{HodgeT4}) indicates the presence of $8$ world-sheet bosons
and $8$ world-sheet fermions in the second quantized symmetric product theory. We will encounter a small subtlety 
in the naive implementation of this idea.

The $SO(9)$ covariance of the flat space partition function was rendered manifest in the Neveu-Schwarz
formalism in \cite{Hanany:2010da}. To exploit these results, it turns out to be useful
to first
translate them into the Green-Schwarz
formalism. The latter more directly connects to the generating function of the chiral
ring of the symmetric product of the four-torus. 
The translation involves the use of the generalized Jacobi identity for
theta-functions, in turn closely related to triality of $SO(8)$. We
briefly review the necessary background.

\subsubsection{The  $SO(8)$ covariant generating function and triality}
Before implementing $SO(9)$ covariance, we manipulate the $SO(8)$ covariant partition function
of flat space superstring theory to switch from the Neveu-Schwarz-Ramond to the Green-Schwarz representation.
The worldsheet bosonic $SO(8)$ covariant factor in the  partition function is:
\begin{eqnarray*}
Z_B  &=& \prod\limits_{l=1}^{\infty} \prod\limits_{\lambda \in
\mathbf{8_v}} \frac{1}{1- y^\lambda q^{l}}=q^{1/3} \eta^4 \prod\limits_{a=1}^4 \frac{2 \sin (\pi \nu_a) }{\theta_1 (\nu_a, \tau)}
\, ,
\end{eqnarray*}
where $\mathbf{8_V}$ is the eight-dimensional vector representation  of $SO(8)$ whose character is denoted $[1,0,0,0]_8$, see the notation (\ref{DynkinNotation}). We have defined the $SO(8)$ fugacities:
\begin{equation}
  e^{2 \pi i \nu_1} = y_1, \quad 
  e^{2 \pi i \nu_2} = \frac{y_2}{y_1} , \quad 
  e^{2 \pi i \nu_3} = \frac{y_3 y_4}{y_2}, \quad 
  e^{2 \pi i \nu_4} = \frac{y_4}{y_3} \, .
\label{so8fugacities}
\end{equation}
For the world sheet fermions we have partition functions that are given by (possibly shifted versions of):
\begin{eqnarray*}
Z_F &=& \prod\limits_{l=0}^{\infty} \prod\limits_{\lambda \in \mathbf{8_v}} (1+f  y^\lambda q^{l}) . 
\end{eqnarray*}
For a given state, the parity of the power of $f$ is equal to  the world sheet fermion number.  With this in
mind, the GSO projection in the NS sector $(-1)^F=1$ consists in
keeping the terms in the partition function that are even powers of
$f$ (after taking into account the fermion number of the covariant vacuum). 
In the Ramond sector, we keep both sectors from the perspective
of the oscillators $(-1)^F=\pm 1$, and associate each
 to different chirality vacuum states (spinor $\mathbf{8_s}$ or
$\mathbf{8_c}$). The total partition function for the left-movers,
taking bosons and fermions into account, is then
\begin{eqnarray}
Z_L &=& \frac{1}{2} ([0,0,1,0]_8  - [0,0,0,1]_8) \\
& & \mkern-50mu +  8\prod\limits_{a=1}^4 \frac{ \sin (\pi \nu_a) }{\theta_1 (\nu_a, \tau)} \left[
 \prod\limits_{a=1}^4  \theta_3 (\nu_a,\tau) -  
\prod\limits_{a=1}^4  \theta_4 (\nu_a,\tau) + \frac{1}{16 } ([0,0,0,1]_8  +[0,0,1,0]_8) \prod\limits_{a=1}^4
\frac{\theta_2 (\nu_a,\tau)}{\cos (\pi \nu_a)} \right] \, .  \nonumber
\end{eqnarray}
Using the identities between characters and trigonometric functions
\begin{equation}
 [0,0,0,1]_8 = 8 \left( \prod\limits_{a=1}^4 \cos (\pi \nu_a) + \prod\limits_{a=1}^4 \sin (\pi \nu_a) \right)
\end{equation}
\begin{equation}
 [0,0,1,0]_8 = 8 \left( \prod\limits_{a=1}^4 \cos (\pi \nu_a) - \prod\limits_{a=1}^4 \sin (\pi \nu_a)
\right) \, ,
\end{equation}  
we can write this in a more symmetric form: 
\begin{equation}
Z_L = 8\prod\limits_{a=1}^4 \frac{ \sin (\pi \nu_a) }{\theta_1 (\nu_a, \tau)} \left[
 \prod\limits_{a=1}^4  \theta_3 (\nu_a,\tau) -  
\prod\limits_{a=1}^4  \theta_4 (\nu_a,\tau) + \prod\limits_{a=1}^4
\theta_2 (\nu_a,\tau) -\prod\limits_{a=1}^4
\theta_1 (\nu_a,\tau) \right]. 
\end{equation}
These are the Neveu-Schwarz-Ramond expressions which were the starting
point of \cite{Hanany:2010da}. The elliptic genus generating function
however is more akin to a Green-Schwarz formalism partition function.  We
therefore wish to translate these $SO(8)$ covariant partition sums
into the Green-Schwarz language, in which bosonic oscillators on the world sheet
form a vector of
$SO(8)$ in light-cone gauge, and fermionic oscillators a spinorial
representation \cite{GSW1}. It is well-known that we can implement the change
of formalism through triality.
We can act with triality on the simple roots of $SO(8)$, e.g. 
exchanging the first and fourth root. 
The action induces an action on the fugacities (see also equation (\ref{so8fugacities}))
\begin{equation}
  e^{2 \pi i \nu_1'} = y_4 \quad 
  e^{2 \pi i \nu_2'} = \frac{y_2}{y_4}\quad 
  e^{2 \pi i \nu_3'} = \frac{y_1 y_3}{y_2}\quad 
  e^{2 \pi i \nu_4'} = \frac{y_1}{y_3} \, .
\end{equation}
This change of variables lifts to the full oscillator spectrum in the
 generalized Jacobi identity \cite{WhitWat}
\begin{eqnarray}
\prod\limits_{a=1}^4 \theta_3 (\nu_a , \tau)  - \prod\limits_{a=1}^4 \theta_4 (\nu_a , \tau) 
+  \prod\limits_{a=1}^4 \theta_2 (\nu_a ,
\tau)-  \prod\limits_{a=1}^4 \theta_1 (\nu_a , \tau) 
&=& 2 \prod\limits_{a=1}^4 \theta_1 (\nu'_a + \frac{1}{2} , \tau) \nonumber
\\ &=& 2 \prod\limits_{a=1}^4 \theta_2 (\nu'_a,
\tau) \, .
\end{eqnarray}
Using this theta function identity in the  partition function, we obtain
\begin{equation}
Z_L =
16 \prod\limits_{a=1}^4 \frac{ \sin (\pi \nu_a) }{\theta_1 (\nu_a, \tau)} 
\theta_2 (\nu'_a,
\tau) . 
\end{equation}
We can rewrite the $\theta$-functions in terms of products, and
use the identities
\begin{equation}
 [1,0,0,0]_8 = 8 \left( \prod\limits_{a=1}^4 \cos (\pi \nu'_a) + \prod\limits_{a=1}^4 \sin (\pi \nu'_a)
\right)
\end{equation}
\begin{equation}
 [0,0,1,0]_8 = 8 \left( \prod\limits_{a=1}^4 \cos (\pi \nu'_a) - \prod\limits_{a=1}^4 \sin (\pi \nu'_a)
\right), 
\end{equation}
to find: 
\begin{equation}
\label{partfunc}
Z_L =  ([1,0,0,0]_8+[0,0,1,0]_8 )
 \prod\limits_{a=1}^4 \prod\limits_{n=1}^{\infty}
\frac{(1+e^{2i \pi \nu'_a}
q^n)(1+e^{-2i \pi \nu'_a} q^n)}{(1-e^{2i \pi \nu_a}
q^n)(1-e^{-2i \pi \nu_a} q^n)}. 
\end{equation}
We recognize this as the Green-Schwarz formalism partition function corresponding to
a $8_V$ vector worth of bosonic oscillators and a $8_S$ spinor of fermionic worldsheet oscillators. The factorized term $[1,0,0,0]_8+[0,0,1,0]_8$ is the 16-dimensional multiplet of massless ground states. 
When we take all fugacities to 1, the partition function can be factorized as 
\begin{equation}
\label{factorizedPartFunc}
Z_L = 16 + 256 \left( q + 9 q^2 + 60 q^3 + 329 q^4 + 1566 q^5 + ...\right) \, .
\end{equation}
Finally, we can address the subtlety that we encounter in identifying the flat
space partition function with the chiral ring generating function. There are $16$ chiral ring
states for a single copy $N=1$ of $T^4$ (corresponding to the term $q^1$ in the generating function (\ref{CFTpartfunc})),
but there are $256$ chiral massive states at the first level in the superstring as seen in (\ref{factorizedPartFunc}). It should also be
noted that the smallest non-trivial representations of $SO(5)$ are of dimensions $4$ and $5$, while
the border of the Hodge diamond (\ref{HodgeDiamondT4}) are representations of dimension $2$ that are, in some cases, non-trivial representations of $SO(4) \subset SO(5)$, according to the discussion below (\ref{HodgeT4rep}). 

We must therefore further specify that the arguments of \cite{deBoer:2008ss} apply to the non-trivially interacting part of the 
boundary superconformal field theory, which we obtain by factoring the center of mass $T^4$ from the symmetric
product conformal field theory. Correspondingly, in the flat space chiral generating function for the massive modes, 
we will strip off the massive supermultiplet degrees of freedom. The minimal requirement of the matching of the number
of degrees of freedom is then met. This can be seen by factoring out a factor of sixteen from equation
(\ref{CFTpartfunc}) and comparing to the term between parentheses in equation (\ref{factorizedPartFunc}).
In the following, we demonstrate that this heedful observation also allows us to implement $SO(5)$ covariance 
in the chiral ring generating function.

\subsubsection{The $SO(9)$ and $SO(5)$ covariant generating functions}

We recall that in \cite{Hanany:2010da} it was demonstrated 
that the $SO(9)$ massive representation content of the
superstring can be recuperated from the formula for the partition function in the
following way: one splits off the massless representation and substitutes $SO(9)$ fugacities for
$SO(8)$ fugacities using the substitution
\begin{equation}
y_3 = z_3/z_4 \, \quad \mbox{and} \quad y_i = z_i \quad \text{for} \,  i=1,2,4 \, ,
\end{equation}
expands, and decomposes. See  \cite{Hanany:2010da} for details. 
It was also shown that one can factor out
a massive supermultiplet factor.
Applying this procedure to the partition function (\ref{partfunc}) gives 
  \begin{equation}
 \label{factorizedPartFunc2}
Z_L = Z_{\mathrm{massless}}+ Z_{\mathrm{massive}} \times Z_{\text{fact}}
 \end{equation}
 where we expand the factorized partition function as
 \begin{equation}
Z_{\text{fact}} =  \sum\limits_{N=1}^{\infty} Q_N q^N \, . 
 \end{equation}
 The first term 
 \begin{equation}
 Z_{\mathrm{massless}} = [1,0,0,0]_8+[0,0,1,0]_8 
 \end{equation}
 in (\ref{factorizedPartFunc2}) accounts for the massless multiplet and is only a $SO(8)$ character, while $Z_{\mathrm{massive}}$ is the universal massive multiplet factor, given in term of $SO(9)$ characters by 
\begin{eqnarray}
Z_{\mathrm{massive}} &=& [2,0,0,0]_9+[1,0,0,1]_9+[0,0,1,0]_9 \, .
\end{eqnarray}
This term corresponds to the factor 256 in (\ref{factorizedPartFunc}). 
Our main focus is the remaining factor $Z_{\text{fact}}$. Inspired by string-string duality,
we generate the $SO(5)$ covariant chiral ring partition function
by breaking $SO(9) \longrightarrow U(1) \times U(1) \times SO(5)$ in the appropriate way.
This is entirely analogous to what we did in subsection \ref{covK3}. We make
the substitutions:
\begin{equation}
q \rightarrow t \bar{t} q
\quad
 z_1 \longrightarrow t \bar{t}  \qquad z_2 
\longrightarrow t^2 \qquad z_3 \longrightarrow t^2 \tilde{z}_1 \qquad z_4
\longrightarrow t \tilde{z}_2 \, .
\end{equation}
For the massive supermultiplet, we obtain:
\begin{eqnarray} \label{Zmasu}
(t \bar{t})^2 Z_{\mathrm{massive}}  = 
\begin{array}{ccc}
 & & 1 \\
 & [01] &  \\
{[}10{]}+1 &  & [02]+[10]+1   \\
 &   [11]+2 \cdot [01] &  \\
  &   & [20]+[02]+2 \cdot [10]+2
\end{array} \, 
\end{eqnarray}
In the final expression, the coefficient of the monomial $t^p \bar{t}^q$ is the character
of the $SO(5)$ representation of the chiral primaries of charge $(p,q)$.
To summarize,  expand (\ref{partfunc}), 
using the characters of the vector, spinor and pseudo-spinor representations
of $SO(8)$, and take the coefficient of $q^N$.
Multiply this coefficient by $(t \bar{t})^{N-1}$ and make the substitution 
\begin{equation}
y_1 \longrightarrow t \bar{t} \qquad 
y_2 \longrightarrow t^2 \qquad 
y_3 \longrightarrow t \frac{\tilde{z}_1}{\tilde{z}_2}\qquad 
y_4 \longrightarrow t \tilde{z}_2 \, .
\end{equation}
 The decomposition of the Hodge diamond into representations of $SO(5)$ is obtained via this
 procedure.
 We use Dynkin label notation for the $SO(5)$ representations, and we arrange the matrices
such that the $U(1) \times U(1)$ quantum numbers act as row and column labels. 
As argued above, we strip off this factor, and continue to study the remaining factor, which corresponds to the 
Hodge diamond of the orbifold conformal field theory where we factor out the center of mass degrees of freedom.
The factor $Z_{\text{fact}}$ is coded in the diamonds $Q_N$
which at levels $N=1,...,5$ read, depicting only one eighth of the diamond following the convention (\ref{8thdiamond}):
\begin{equation} 
Q_1= 1 
\end{equation}
\begin{equation} 
Q_2=
\begin{array}{ccccc}
 & & 1 & & \\
  & 0 & & 0 & \\
  1 & & [10] & & 1 \\
  & 0 & & 0 & \\
  & & 1 & &
\end{array}
= 
\begin{array}{cc}
 & 1 \\
 0 & \\
 & [10] 
\end{array}
\end{equation}
\begin{equation} 
Q_3=
\begin{array}{ccc}
 & & 1 \\
 &0  & \\
 1 & & [10] \\
 &[01]  & \\
  \hdashline 
 & & [20]+2 
\end{array}
\end{equation}
\begin{equation} \label{Q4}
Q_4=
\begin{array}{cccc}
 & & & 1 \\
 & & 0 & \\
 & 1 & & [10] \\
 0 & & [01] & \\
 & [10]+1 & & [20]+[10]+3 \\
  \hdashline 
 & & [11]+2[01] & \\
 & & & [30]+[02]+3[10]+2 
\end{array}
\end{equation}
\begin{equation}
Q_5=
\begin{array}{ccccc}
 & & & & 1  \\
& & & 0 &  \\
& & 1 &  & [10] \\
 & 0 & & [01] &  \\
1  & & [10]+1 & & [20]+[10]+3  \\ 
 & [01]  & & [11]+3[01] &   \\
 \hdashline 
  & & [20]+3[10]+4 & & [30]+[20]+2[02]+4[10]+4  \\
   & & & [21]+3[11]+6[01] &   \\
    & & & &   [40]+[12]+3[20]+2[02]+6[10]+8
\end{array}
\end{equation}
We dropped the index $5$ for the characters -- they are all associated to $SO(5)$. We note that the $(i,j)$th entry of $Q_{N}$ is independent of $N$ provided $N \geq (i-1)+(j-1)$. Those entries are above the dashed line in the diamonds. When this condition is fulfilled, the dimensions of the representations match those of the reduced Hodge diamond (\ref{reducedHodgeT4}) of the tensor product theory. For instance, assume we want to know the decomposition of the representation of dimension $167$ in (\ref{reducedHodgeT4}), which corresponds to the entry $(6,4)$ in the matrices $Q_N$ for any $N \geq 8$. The associated character can be computed as outlined above, and one finds for the $(6,4)$ coefficient of $Q_8$:
\begin{equation}
\chi_{167} = [30] + 3[20] + 4[02] + 9[10] + 10 \, . 
\end{equation}
These decompositions are obtained by the identification of characters in the covariantized generating function.

\subsubsection{Summary}
We related the $SO(5)$ covariance
of the chiral ring of the superconformal dual to $AdS_3 \times S^3
\times T^4$ to a broken Lorentz little group for chiral massive type
II superstring excitations. Again, the inspiration lies in
string-string dualities that relate the chiral primaries to half BPS states \cite{Dabholkar:1990yf}
in asymptotically flat space.

\section{Conclusions}
\label{threepoint}
We have provided a generating function that codes all of the
$SO(21)$ or $SO(5)$ representation theoretic content of the chiral
ring. We have already reviewed in the introduction how this information is sufficient to compute the connection
on the homogeneous vector bundle spanned by the chiral states over the moduli space \cite{deBoer:2008ss}.
It would be interesting to exploit this knowledge to analyze the $SO(21)$ covariance of known correlation
functions. A much more ambitious goal would be to characterize $1/N$ corrections in terms of $SO(21)$ representation
theoretic data. 

To exploit the covariance fully, it is necessary to identify  the states that fill out the various $SO(21)$
or $SO(5)$ multiplets at a given point in the moduli space. 
Although we did this in the oscillator formalism, one needs to carefully match these states
onto those of the orbifold symmetric product conformal field theory, and in this identification one must allow for operator mixing. Moreover, the operator mixing may involve mixing of single and multi-particle states.
Still, it is clear that if one is able to overcome these technical challenges, the $SO(21)$ classification of the spectrum will become powerful in organizing the dynamical information of the theory, and for instance in predicting new correlation functions from old ones, by completing $SO(21)$ or $SO(5)$ multiplets. We leave this interesting task of mixing and separation for future work.

\section*{Acknowledgments}
 We would like to acknowledge support from the grant ANR-13-BS05-0001, and from the \'Ecole Polytechnique and the \'Ecole Nationale Sup\'erieure des Mines de Paris.

\appendix

\section{Representations and characters}
\label{representationtheory}
In this appendix, we gather some results on the characters and the representation
theory of classical Lie algebras and apply them to the $\mathfrak{so}(n)$ algebras appearing in the body
of the paper.
\subsection{Characters}
Each unitary irreducible representation of a classical Lie algebra is characterized
by its highest weight $\lambda$. The character associated to the representation
with highest weight $\lambda$ is defined as 
\begin{equation}
 \chi_{\lambda}=\sum\limits_{\lambda' \in \Lambda_{\lambda}} \mathrm{mult}_{\Lambda}(\lambda ') e^{\lambda '}
 \, ,
 \label{characterformula}
\end{equation}
where the sum is over all weights $\lambda '$ in the weight system
 $ \Lambda_{\lambda}$ of the representation of highest weight $\lambda$. 
A useful basis of the  lattice of weights is given by the fundamental weights $(\pi_1,...,\pi_r)$.
They have the property that they are orthonormal to a basis of simple co-roots with respect to a natural
non-degenerate form on the classical Lie algebra. Their number is equal to the rank $r$ of the Lie
algebra. The expansion coefficients of a highest weight in the fundamental weight basis are called
Dynkin labels. They can be obtained by computing the inner product of the simple co-roots with the 
weights. Decomposing the weights
\begin{equation}
 \lambda' =
\sum\limits_{i=1}^{r} \lambda'_i \pi_i,
\end{equation}
in terms of $r$ integers $\lambda_i'$,
the formula (\ref{characterformula}) for the characters becomes 
\begin{equation}
 \chi_{\lambda}=\sum\limits_{\lambda' \in \Lambda_{\lambda}} \mathrm{mult}_{\Lambda}(\lambda ')
\prod\limits_{i=1}^{r} e^{\lambda'_i \pi_i} \, .
\end{equation}
Let's define fugacities $y_i=e^{\pi_i}$. Then the characters can be written as a sum of
monomials:
\begin{equation}
 \chi_{\lambda}=\sum\limits_{\lambda' \in \Lambda_{\lambda}} \mathrm{mult}_{\Lambda}(\lambda ')
\prod\limits_{i=1}^{r} {y_i}^{\lambda'_i} \, .
\end{equation}
We use square brackets around Dynkin labels to denote the character associated to a given weight: 
\begin{equation}
\label{DynkinNotation}
[\lambda_1 , \dots \lambda_r]_n = \chi_\lambda \qquad \text{where} \qquad \lambda = \sum\limits_{i=1}^{r} \lambda_i \pi_i  \, . 
\end{equation}
In the text we discuss only characters of algebras of type $\mathfrak{so}$, and we add a subscript $n$ to indicate that we consider the character of an $\mathfrak{so}(n)$ representation. 
Given the character of a reducible representation, one can decompose
it into irreducible representations by identifying the highest weight
in the reducible representation, subtracting the character of the  corresponding
irreducible representation, and continuing recursively. For finite dimensional representations,
this is a finite algorithm that can be implemented on a computer.

\subsection{The $\mathfrak{so}(2r+1)$ and $\mathfrak{so}(2r)$ algebras}
We will have particular use for the Lie algebras associated to the orthogonal groups, namely
$B_r = \mathfrak{so}(2r+1)$ and $D_r = \mathfrak{so}(2r)$.
Take an orthonormal basis $(\epsilon_1,...,\epsilon_r)$ in the space of weights. 
The simple roots and the fundamental weights of $B_r$ can be parameterized as 
\begin{eqnarray}
 \alpha_i &=&  \epsilon_i-\epsilon_{i+1} \qquad \text{for} \, \, \,  \forall  \, i<r \nonumber \\
\alpha_r &=& \epsilon_r  \, ,
\end{eqnarray}
and
\begin{eqnarray}
  \pi_i &=& \epsilon_1 + ... + \epsilon_i \qquad  \text{for} \, \, \, i < r \nonumber \\
  \pi_r &=& 
  (\epsilon_1 + ... + \epsilon_r)/2  \, ,
\end{eqnarray}
while for the algebra $D_r$ we have
\begin{eqnarray}
\alpha_i &=& \epsilon_i-\epsilon_{i+1} \qquad \text{for} \, \, \, i<r \nonumber \\
\alpha_r &=& \epsilon_{r-1}+\epsilon_r \, ,
\end{eqnarray}
and
\begin{eqnarray}
\pi_i &=& \epsilon_1 + ... + \epsilon_i \qquad  \text{for} \, \, \, i < r-1 \nonumber \\
\pi_{r-1} &=& (\epsilon_1 + ... + \epsilon_{r-1} - \epsilon_r)/2 \nonumber \\
\pi_{r} &=& (\epsilon_1 + ... + \epsilon_{r-1} + \epsilon_r)/2  \, .
\end{eqnarray}
We can relate the fundamental weights of these algebras of equal rank through the relations 
\begin{eqnarray}
  \pi^B_i &=& \pi^D_i  \qquad \text{for} \, \, \,  1 \leq i \leq r, \, i \neq r-1 \nonumber \\
  \pi^B_{r-1} &=&
\pi^D_{r-1}+\pi^D_r \, .  
\label{eq102}
\end{eqnarray}
Let us define the fugacities $y_i=\exp(\pi^D_i)$ and $z_i=\exp(\pi^B_i)$. 
Then the relations between the fundamental weights (\ref{eq102}) translate into relations
between the fugacities
\begin{equation}
 \left\{ \begin{array}{ll}
  z_i = y_i & \forall  i \neq r-1 \\
  z_{r-1} =
y_{r-1}y_r & 
 \end{array} \right.
\Leftrightarrow 
 \left\{ \begin{array}{ll}
  y_i = z_i & \forall i \neq r-1 \\
  y_{r-1} =
z_{r-1}/z_r & 
 \end{array} \right.
\label{substitution}
\end{equation}
The important result is then that {\em if} $\chi^D$ is the character of a representation
of $D_r$, which is also a representation of the algebra $B_r$,
 {\em then} the latter can be obtained from the former through the substitution
of the fugacities $y_i$ by the fugacities $z_j$.


\begin{thebibliography}{99}

\bibitem{Maldacena:1997re}
  J.~M.~Maldacena,
  ``The Large N limit of superconformal field theories and supergravity,''
  Adv.\ Theor.\ Math.\ Phys.\  {\bf 2} (1998) 231
  [hep-th/9711200].
  
\bibitem{Strominger:1996sh}
  A.~Strominger and C.~Vafa,
  ``Microscopic origin of the Bekenstein-Hawking entropy,''
  Phys.\ Lett.\ B {\bf 379} (1996) 99
  [hep-th/9601029].
  
\bibitem{Strominger:1997eq}
  A.~Strominger,
  ``Black hole entropy from near horizon microstates,''
  JHEP {\bf 9802} (1998) 009
  [hep-th/9712251].
  
\bibitem{Gaberdiel:2007vu}
  M.~R.~Gaberdiel and I.~Kirsch,
  ``Worldsheet correlators in AdS(3)/CFT(2),''
  JHEP {\bf 0704} (2007) 050
  [hep-th/0703001].
  
\bibitem{Dabholkar:2007ey}
  A.~Dabholkar and A.~Pakman,
``Exact chiral ring of AdS(3) / CFT(2),''
  Adv.\ Theor.\ Math.\ Phys.\  {\bf 13} (2009) 409
  [hep-th/0703022].
  
\bibitem{Jevicki:1998bm}
  A.~Jevicki, M.~Mihailescu and S.~Ramgoolam,
  ``Gravity from CFT on S**N(X): Symmetries and interactions,''
  Nucl.\ Phys.\ B {\bf 577} (2000) 47
  [hep-th/9907144].
  
\bibitem{Lunin:2001pw}
  O.~Lunin and S.~D.~Mathur,
  ``Three point functions for M(N) / S(N) orbifolds with N=4 supersymmetry,''
  Commun.\ Math.\ Phys.\  {\bf 227} (2002) 385
  [hep-th/0103169].
  
\bibitem{Troost:2011ud}
  J.~Troost,
  ``The $AdS_3$ central charge in string theory,''
  Phys.\ Lett.\ B {\bf 705} (2011) 260
  [arXiv:1109.1923 [hep-th]].
  
\bibitem{deBoer:2008ss}
  J.~de Boer, J.~Manschot, K.~Papadodimas and E.~Verlinde,
  ``The Chiral ring of AdS(3)/CFT(2) and the attractor mechanism,''
  JHEP {\bf 0903} (2009) 030
  [arXiv:0809.0507 [hep-th]].
  
\bibitem{Seiberg:1999xz}
  N.~Seiberg and E.~Witten,
  ``The D1 / D5 system and singular CFT,''
  JHEP {\bf 9904} (1999) 017
  [hep-th/9903224].
  
\bibitem{Dijkgraaf:1998gf}
  R.~Dijkgraaf,
``Instanton strings and hyperKahler geometry,''
  Nucl.\ Phys.\ B {\bf 543} (1999) 545
  [hep-th/9810210].
  
\bibitem{Dijkgraaf:1996xw}
  R.~Dijkgraaf, G.~W.~Moore, E.~P.~Verlinde and H.~L.~Verlinde,
  ``Elliptic genera of symmetric products and second quantized strings,''
  Commun.\ Math.\ Phys.\  {\bf 185} (1997) 197
  [hep-th/9608096].
  
\bibitem{Vafa:1994tf}
  C.~Vafa and E.~Witten,
 ``A Strong coupling test of S duality,''
  Nucl.\ Phys.\ B {\bf 431} (1994) 3
  [hep-th/9408074].
  
\bibitem{Dabholkar:1990yf}
  A.~Dabholkar, G.~W.~Gibbons, J.~A.~Harvey and F.~Ruiz Ruiz,
 ``Superstrings and Solitons,''
  Nucl.\ Phys.\ B {\bf 340} (1990) 33.
  
  
  
\bibitem{GSW1}
  M.~B.~Green, J.~H.~Schwarz and E.~Witten,
  ``Superstring Theory. Vol. 1: Introduction,''
   Cambridge Monogr. Math. Phys.
   
\bibitem{Okuyama:2008th}
  K.~Okuyama,
  ``N=4 SYM on K3 and the AdS(3)/CFT(2) Correspondence,''
  JHEP {\bf 0802} (2008) 036
  [arXiv:0801.1313 [hep-th]].
  
\bibitem{Hanany:2010da}
  A.~Hanany, D.~Forcella and J.~Troost,
  ``The Covariant perturbative string spectrum,''
  Nucl.\ Phys.\ B {\bf 846} (2011) 212
  [arXiv:1007.2622 [hep-th]].

\bibitem{WhitWat}
  E.~T.~Whittaker, 
    G.~N.~Watson,
A Course of Modern Analysis,
4th Edition, Cambridge Mathematical Library, 1996.
 \end{thebibliography}
\end{document}